\documentclass[conference,9pt]{IEEEtran}
\IEEEoverridecommandlockouts
\IEEEoverridecommandlockouts
\usepackage{booktabs} 
\usepackage{amsmath}
\usepackage{amsmath,amssymb,amsfonts}
\usepackage{algorithm}
\usepackage{algpseudocode}
\usepackage{algorithmicx}
\usepackage{booktabs}
\usepackage{graphicx}
\usepackage{comment}
\usepackage{multirow}
\usepackage{subcaption}
\usepackage{comment}
\usepackage{hyperref}
\usepackage{hyperref}
\usepackage{booktabs, dcolumn}
\usepackage{soul}
\usepackage{dcolumn}
\usepackage{enumitem}
\usepackage{cite}
\usepackage{pifont}
\usepackage[detect-none]{siunitx}
\usepackage[export]{adjustbox}
\usepackage{mathtools}

\makeatletter
\newcommand{\vast}{\bBigg@{4}}
\newcommand{\Vast}{\bBigg@{5}}
\makeatother

\usepackage[skip=2pt]{caption} % example skip set to 2pt
\usepackage[switch]{lineno}

\renewcommand{\IEEEbibitemsep}{0pt plus 0.5pt}
\makeatletter
\IEEEtriggercmd{\reset@font\normalfont\fontsize{7.5pt}{8.40pt}\selectfont}
\makeatother
\IEEEtriggeratref{1}

\usepackage{lipsum} 
\usepackage{xargs} 
\usepackage[pdftex,dvipsnames]{xcolor}  
\usepackage[colorinlistoftodos,prependcaption]{todonotes}
\newcommandx{\unsure}[2][1=]{\todo[linecolor=red,backgroundcolor=red!25,bordercolor=red,#1]{#2}}
\newcommandx{\change}[2][1=]{\todo[linecolor=blue,backgroundcolor=blue!25,bordercolor=blue,#1]{#2}}
\newcommandx{\info}[2][1=]{\todo[linecolor=OliveGreen,backgroundcolor=OliveGreen!25,bordercolor=OliveGreen,#1]{#2}}
\newcommandx{\improvement}[2][1=]{\todo[linecolor=Plum,backgroundcolor=Plum!25,bordercolor=Plum,#1]{#2}}
\newcommandx{\thiswillnotshow}[2][1=]{\todo[disable,#1]{#2}}
\usepackage{overpic} 
\newcolumntype{d}[1]{D{.}{.}{#1}}

\usepackage[utf8]{inputenc}
\usepackage{listings}
\usepackage{xcolor}

\definecolor{codegreen}{rgb}{0,0.6,0}
\definecolor{codegray}{rgb}{0.5,0.5,0.5}
\definecolor{codepurple}{rgb}{0.58,0,0.82}
\definecolor{backcolour}{rgb}{0.95,0.95,0.92}
\lstdefinestyle{mystyle}{
  backgroundcolor=\color{white},
  xleftmargin = 0.2cm,
  commentstyle=\color{codegreen},
  keywordstyle=\color{magenta},
  numberstyle=\footnotesize\color{codegray},
  stringstyle=\color{codepurple},
  basicstyle=\ttfamily\footnotesize,
  breakatwhitespace=false,         
  breaklines=true,                 
  captionpos=b,                    
  keepspaces=false,                 
  numbers=left,                    
  numbersep=5pt,                  
  showspaces=false,                
  showstringspaces=false,
  showtabs=false,                  
  tabsize=2, 
  lineskip=-2ex, % <---
  escapechar={|}
}
\usepackage{tikz}
\usetikzlibrary{decorations.pathreplacing,calc,shapes,positioning,tikzmark}

\newcommand{\blue}[1]{\textcolor{black}{#1}}

\IEEEoverridecommandlockouts
\IEEEpubid{\makebox[\columnwidth]{978-1-6654-3274-0/21/\$31.00~\copyright~2021~IEEE \hfill}
\hspace{\columnsep}\makebox[\columnwidth]{ }}

\begin{document}
%\linenumbers

\title{Skew-Oblivious Data Routing for Data Intensive Applications on FPGAs with HLS}
\author{
\Large Xinyu Chen$^{1}$, Hongshi Tan$^{1}$, Yao Chen$^{2}$, Bingsheng He$^{1}$, Weng-Fai Wong$^{1}$, Deming Chen$^{2,3}$ \vspace{1mm} \\
\large
$^{1}$National University of Singapore,
$^{2}$Advanced Digital Sciences Center,
$^{3}$University of Illinois at Urbana-Champaign \\
\vspace{-8mm}
}

\maketitle
%\IEEEpubidadjcol

\begin{abstract}
FPGAs have become emerging computing infrastructures for accelerating applications in datacenters. 
Meanwhile, high-level synthesis (HLS) tools have been proposed to ease the programming of FPGAs. % and perform well on regular applications. 
%While HLS performs well on regular applications, multiple {processing elements} (PEs) with each owning a private BRAM-based buffer need to be expressed explicitly for irregular data-intensive applications.
\blue{Even with HLS, irregular data-intensive applications require explicit optimizations, among which multiple {processing elements} (PEs) with each owning a private BRAM-based buffer are usually adopted to process multiple data per cycle.
}
%\textcolor{red}{Instead of having inclusive buffers for PEs, the data routing based architecture embraces exclusive buffers, thus increasing the buffered data size.}
Data routing, which dynamically dispatches multiple data to designated PEs, avoids data replication in buffers compared to statically assigning data to PEs, hence saving BRAM usage.
However, the workload imbalance among PEs vastly diminishes performance when processing skew datasets.
%caused by the data skew prohibits the effectiveness of data routing.
In this paper, we propose a skew-oblivious data routing architecture that allocates secondary PEs and schedules them to share the workload of the overloaded PEs at run-time.
In addition, we integrate the proposed architecture into a framework called Ditto to minimize the development efforts for applications that require skew handling.
%assist developers in mapping applications that could benefit from data routing and HLS. 
We evaluate Ditto on five commonly used applications: histogram building, data partitioning, pagerank, heavy hitter detection and hyperloglog.
The results demonstrate that the generated implementations are robust to skew datasets and outperform the state-of-the-art designs in both throughput and BRAM usage efficiency.
\end{abstract}

\IEEEpeerreviewmaketitle

\section{Introduction}
%Due to the failure of Dennard scaling and the appearance of Dark Silicon, successive CPU generations exhibit diminishing performance returns. 
%Benefiting from high performance, energy efficiency and reconfigurability, field-programmable gate arrays (FPGAs) have attracted increasing attention for accelerating datacenter applications. 
%For example, Microsoft adopted FPGAs to accelerate Bing search engine~\cite{Bing_engine}; Baidu introduced FPGAs for machine learning platform~\cite{Ouyang2014}. 

{
Benefiting from high performance, energy efficiency and reconfigurability, field-programmable gate arrays (FPGAs) have attracted increased attention for accelerating datacenter applications.
Meanwhile, in order to provide better programmability for FPGA-based application acceleration, both industry and academia are actively developing high-level synthesis ({HLS}) tools, which can transform {\em kernels} written in high-level description languages to efficient FPGA accelerators without involving tedious and error-prone hardware description language ({HDL}) based programming~\cite{HLSsurvey}. 
}
%In general, HLS extracts parallelism of {regular} applications well by static analysis at compile time~\cite{Li2017}.

{
Still, for {irregular} {data-intensive} applications (e.g., database, graph processing, and in-network processing applications~\cite{ST-Accel}), HLS requires a number of explicit optimizations~\cite{cong2018automated,thomas2020fleet,cong2017bandwidth,Li2017}, which include two important {ones}: 1) on-chip data buffering with BRAMs to alleviate the poor locality caused by irregular memory access patterns; 2) multiple {processing elements} (PEs) with each owing a private buffer to process multiple data per cycle. 
Subsequently, multiple unordered data needs to be dispatched to multiple PEs in one cycle for a balanced pipeline.
A static dispatching scheme (e.g., assigning the $i$-th data to the $i$-th PE) is simple for implementation but requires each PE to keep a replica of the buffered data~\cite{fccmspmv}.
In contrast, dynamically dispatching data to their designated PEs (termed as data routing) enables PE to buffer only a partial range of data, hence saving BRAM usage.
Given that HLS-based data routing has shown superior efficiency on graph processing~\cite{chen2019fly,chen2021thundergp} and database~\cite{chen2020fpga} problems, we believe it is applicable for a class of irregular data-intensive applications.
}

{
Despite the effectiveness of data routing, a largely overlooked problem is the workload imbalance among PEs introduced by data skew. 
Since PEs process distinctive ranges of data, skew datasets may cause some PEs overloaded or underutilized, which essentially diminishes performance.
The challenge of skew handling for data-intensive applications is that the lightweight computation (e.g., the calculation with integers finished within one cycle) cannot tolerate any heavy workload rebalancing operations such as atomic-based work-stealing~\cite{workstealing}.
Besides, skew handling needs to adapt to very different data distributions in a robust manner and requires sizable hardware expertise in general; therefore, the other challenge is to minimize the manual development efforts for developers.
%Besides, skew handling is generally a complex problem that requires sizable hardware expertise; 
In order to address the challenges mentioned above, we propose a skew-oblivious data routing solution with the following key contributions:
}

\begin{itemize}[leftmargin=*]
\item We propose an adaptive skew-oblivious data routing architecture, which allocates secondary PEs that own private buffers and dynamically schedules them to help overloaded PEs at run-time.

\item We integrate the proposed architecture into a framework called \textit{Ditto}\footnote{Ditto, a Transform Pokémon, which is able to reconstitute entire cellular structure to change into what it sees.}, which takes the high-level specification of data-intensive applications as input and outputs the most efficient implementation for the given dataset.

\item We evaluate Ditto on five commonly used data-intensive applications. 
Ditto delivers up to 2.4$\times$ performance speedup and 32$\times$ BRAM usage reduction over state-of-the-art designs on uniform datasets, and outperforms baseline by 12$\times$ on skew datasets.
\end{itemize}

\section{Motivation}\label{sec:motivation}
This section illustrates our motivation of skew handling for data routing with an example -- histogram building (HISTO).

Listing~\ref{list:histo} shows the algorithm of HISTO. 
For each tuple, the destination bin is calculated through a hash function, and the count of the current bin is correspondingly increased by one. 
When implementing HISTO on FPGAs, bins are buffered in on-chip buffers constructed by BRAMs to hide the long latency of bin indexing by the hash value, and multiple PEs are adopted to fully utilize the memory bandwidth. 
In order to provide concurrent read/write accesses~\cite{cong2018automated}, instead of providing a shared buffer pool for all PEs, buffers are partitioned to make every PE own a private buffer.
Assuming the memory interface reads eight tuples per cycle and a PE processes one tuple every two cycles (one cycle for reading the value from and one cycle for writing the result to the buffer), the design requires 16 PEs with 16 buffers.

\begin{minipage}[b]{1\linewidth}
\begin{lstlisting}[language=C++, caption=The code snippet of histogram building.,label={list:histo}]
for(i = 0; i < num_tuples; i ++){ 
    struct {int key, value;} tuple = relation[i];
    int idx = hash(tuple.key);
    Bin[idx] += 1;
}
\end{lstlisting}
\vspace{-3mm}
\end{minipage}
%    |\color{magenta} int2| tuple = relation[i]; // <int key, int value>

\subsection{Benefits of Data Routing}
\begin{figure}[t]
    \centering
    \begin{subfigure}[t]{0.24\textwidth}
        \centering
        %\smallskip
        \includegraphics[width=1\linewidth]{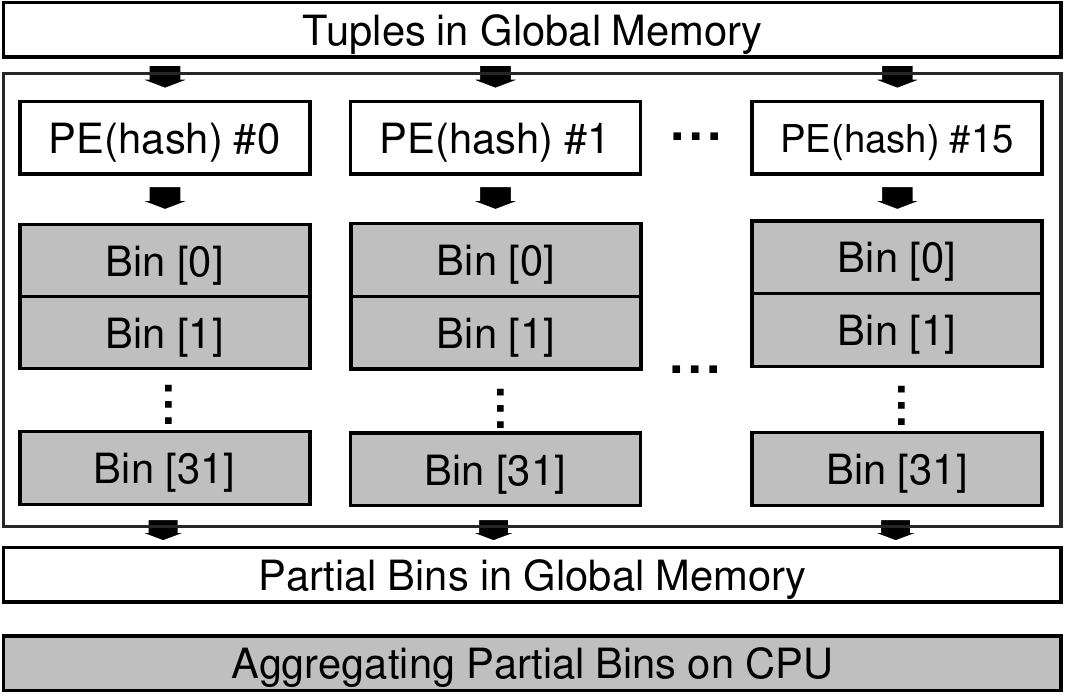}
        \caption{Existing HLS-based HISTO.}
        \label{fig:histo_independentPE}
    \end{subfigure}
    \smallskip
    \hfill
    \begin{subfigure}[t]{0.24\textwidth}
        \centering
        \includegraphics[width=1\textwidth]{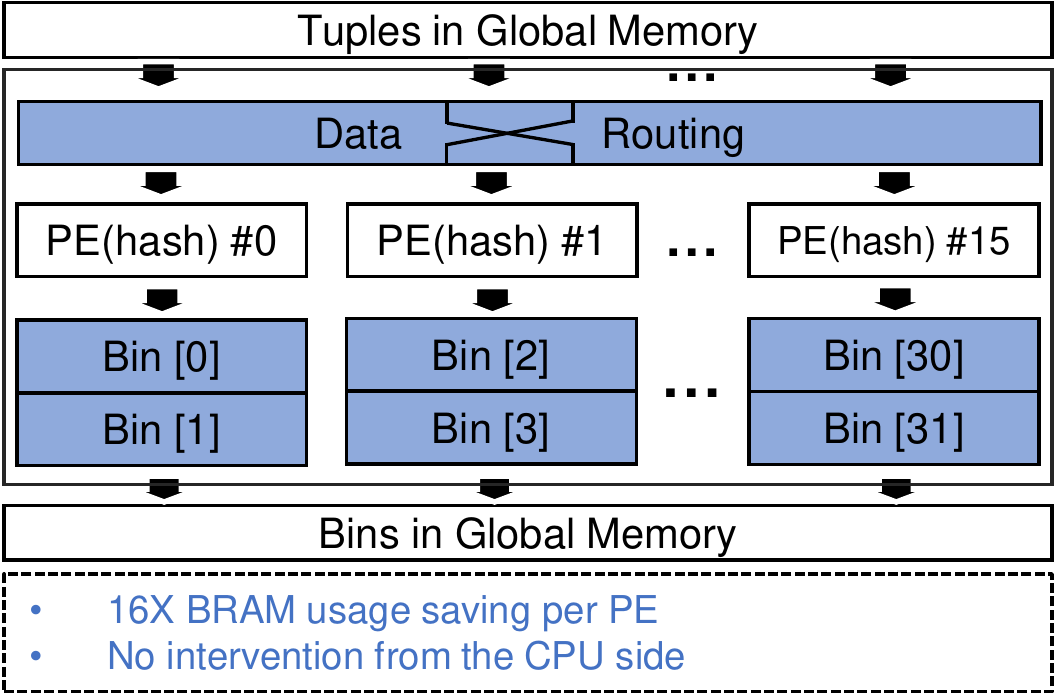}
        \caption{Data routing-based HISTO.}
        \label{fig:histo_datarouting}
    \end{subfigure}
    \vspace{-1.5mm}
    \caption{Designs of HLS-based HISTO with 16 PEs and 32 bins.}
    \label{fig:archs_datapartitioning}   
\end{figure}

There are two bandwidth-optimal schemes for dispatching multiple unordered tuples to multiple PEs. 
Fig.~\ref{fig:histo_independentPE} shows the HISTO with 32 bins from existing HLS-based works~\cite{cong2018automated,jiang2020boyi}, where tuples are statically assigned to PEs and bins are replicated in buffers. Fig.~\ref{fig:histo_datarouting} depicts the data routing based HISTO with 32 bins, where buffers keep distinctive bins and tuples are routed to designated PEs.

The benefit of data routing is twofold.
Firstly, by omitting bin replication, data routing based HISTO saves BRAM usage compared to existing HISTO~\cite{jiang2020boyi}, which is proportional to the number of PEs.
Secondly, existing HISTO requires the intervention of CPU side to aggregate bins for final results; whereas, data routing based HISTO resolves the coordination of PEs and outputs final bins directly without further orchestration. 
%\textcolor{blue}{Similarly, data routing logic shows great potentials for other data-intensive applications.}

\subsection{Workload Imbalance of Data Routing on Skew Datasets}
\begin{figure}[h]
    \centering
    \begin{subfigure}[t]{0.243\textwidth}
        \centering
        \includegraphics[width=1\textwidth]{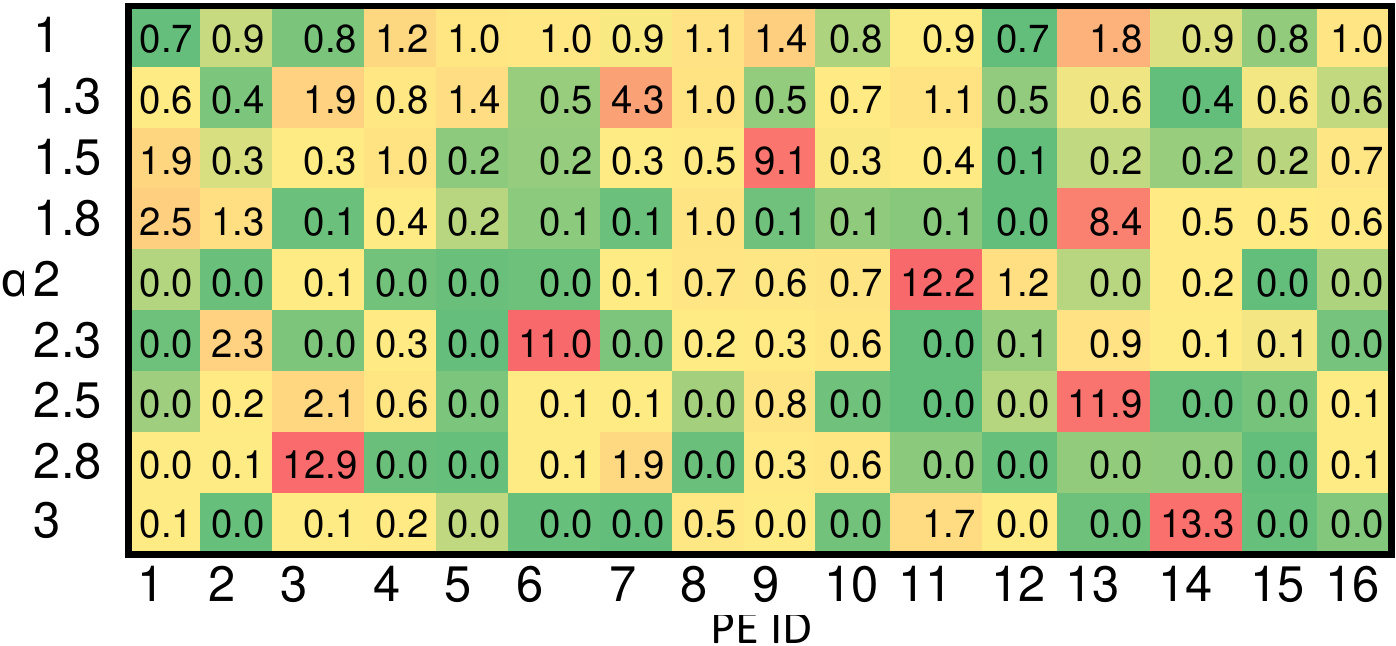}
        \caption{Workload distribution of 16 PEs.}
        \label{fig:workloaddistributionHISTO}
    \end{subfigure}
    \smallskip
    \begin{subfigure}[t]{0.235\textwidth}
        \centering
        \includegraphics[width=1\linewidth]{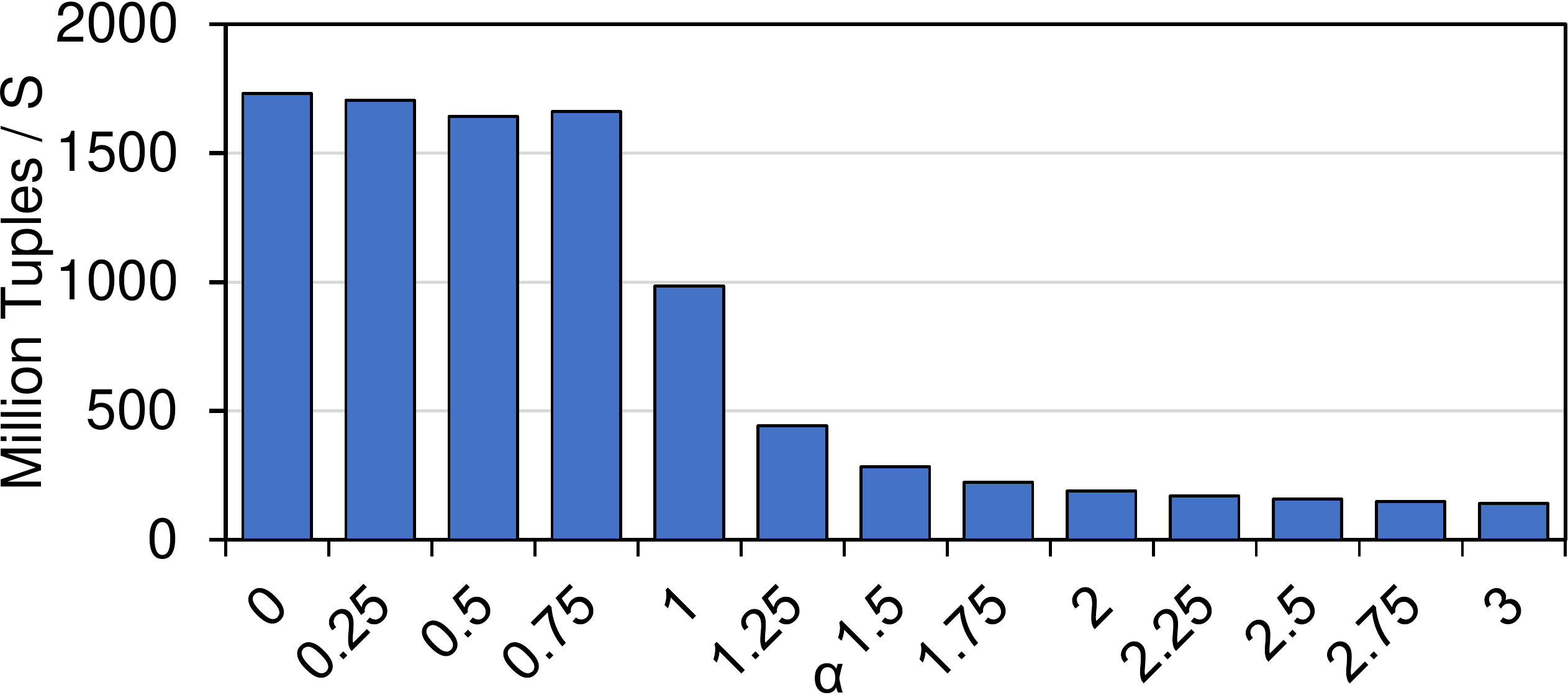}
        \caption{Throughput with varying $\alpha$.}
        \label{fig:ThroughputwithzipfHISTO}
    \end{subfigure}
    \vspace{-1.5mm}
    \caption{Experimental results of HISTO on Zipf datasets.}
    \label{fig:profile_HISTO}   
\end{figure}

As data routing is data-dependent, skew datasets potentially cause workload imbalance among PEs. 
To conduct quantitative analysis, we implement HISTO with 16 PEs on our hardware platform (shown in Section~\ref{sec:hardware}) and profile the execution with 26 million tuples (8-byte) under the Zipf distribution~\cite{Balkesen2013}.
Fig.~\ref{fig:workloaddistributionHISTO} shows the heatmap of the workload (number of tuples) distribution of 16 PEs, {which is normalized to that of the uniformly distributed dataset} ($\alpha = 0$). 
Fig.~\ref{fig:ThroughputwithzipfHISTO} shows the throughput of HISTO with varying the Zipf factor.

As shown in Fig.~\ref{fig:workloaddistributionHISTO}, there is a clear workload imbalance among PEs (red ones are overloaded while green ones are underutilized).
Significant Zipf factor results in severe workload imbalance. 
Besides, overloaded PEs vary across datasets with different Zipf factors.
As a result, as shown in Fig.~\ref{fig:ThroughputwithzipfHISTO}, the throughput of HISTO downgrades significantly with increased Zipf factors.
The performance of the extreme skew dataset ($\alpha = 3$) has slowed down to one-sixteenth of that of the uniform dataset since almost all tuples go to the same PE.
As skew datasets are usual inputs for data-intensive applications in the real world, the workload imbalance problem needs to be well addressed to achieve high performance.

\section{Challenges and Solutions}\label{sec:overview}
While the load-balancing problem has been studied extensively, the data skew introduces unique challenges on data intensive-applications with HLS. 
In this section, we discuss the challenges and our solutions. 

%\subsection{Challenges}\label{sec:chanllenges}
\noindent\textit{\textbf{Challenge 1: How to handle skew with lightweight computation?}}
Data-intensive applications usually have lightweight computation, and the throughput of PEs is bounded by reading data from and writing results to the corresponding buffers~\cite{cong2018automated}.
As a result, underutilized PEs stealing the workload from the overloaded PEs and writing the results back to their buffers after the calculation will not payoff~\cite{awbgcn}.
In addition, heavy operations (e.g., atomic operation) will stall the processing pipeline, resulting in new system bottlenecks~\cite{workstealing}.
%More specially, the buffers are exclusive of each other, which prohibits underutilized PEs from stealing the workload of overloaded PEs~\cite{workstealing}.

%\vspace{1mm}
\noindent\textit{\textbf{Challenge 2: How to minimize manual efforts for skew handling?}}
Skew handling is a complex operation that needs careful hardware optimization. 
On the other hand, different levels of data skew require architectures with different skew handling capacities. To not overshadow the productivity of HLS-based development, data skew should be handled with good programmability.

%Skew handling is beneficial for various data-intensive applications that have different computation patterns;
%hence, generalized skew handling method requires extracting common functionalities of these applications while providing easy-to-use programmability. 
%While the work-stealing is a well-known solution for workload imbalance, it does not pay off on data-intensive applications.
%For work-stealing, the under-utilized PEs steal the tuples from over-loaded PEs and write results back after the calculation.
%\subsection{Solution Overview}\label{sec:oursolution}
\noindent
\textit{\textbf{Solution to challenge 1: A skew-oblivious data routing architecture.}}
As the bottleneck of a PE's throughput relies on the number of ports of the buffer, the principle of skew-oblivious data routing is equivalent to ``increasing the number of buffer ports for overloaded PEs". Instead of having a fixed number of PEs, our solution allocates secondary PEs that own private buffers and dynamically schedules them to share the workload of the overloaded PEs. While more secondary PEs provide more buffer ports, they cost more BRAM resources; hence, the solution introduces a trade-off between the BRAM usage and the skew handling capacity.

%\vspace{1mm}
\noindent\textit{\textbf{Solution to challenge 2: A skew-oblivious framework -- Ditto.}}
We propose Ditto, a framework integrated with the proposed skew-oblivious data routing architecture to ease skew handling for data-intensive applications.
With Ditto, developers only need to write high-level specifications without touching hardware design details, and Ditto generates the most suitable hardware implementation that handles data skew in a given dataset and optimizes the BRAM usage.

\section{Skew-oblivious Data Routing Architecture}\label{sec:arch_templates}
This section describes the proposed skew-oblivious data routing architecture, which resolves the workload imbalance problem caused by skew datasets of the previous HLS-based data routing logic~\cite{chen2019fly}.
%\vspace*{-0.3em}
\subsection{Architecture Overview}
\begin{figure}[h]
    \centering
    \includegraphics[width=1\linewidth]{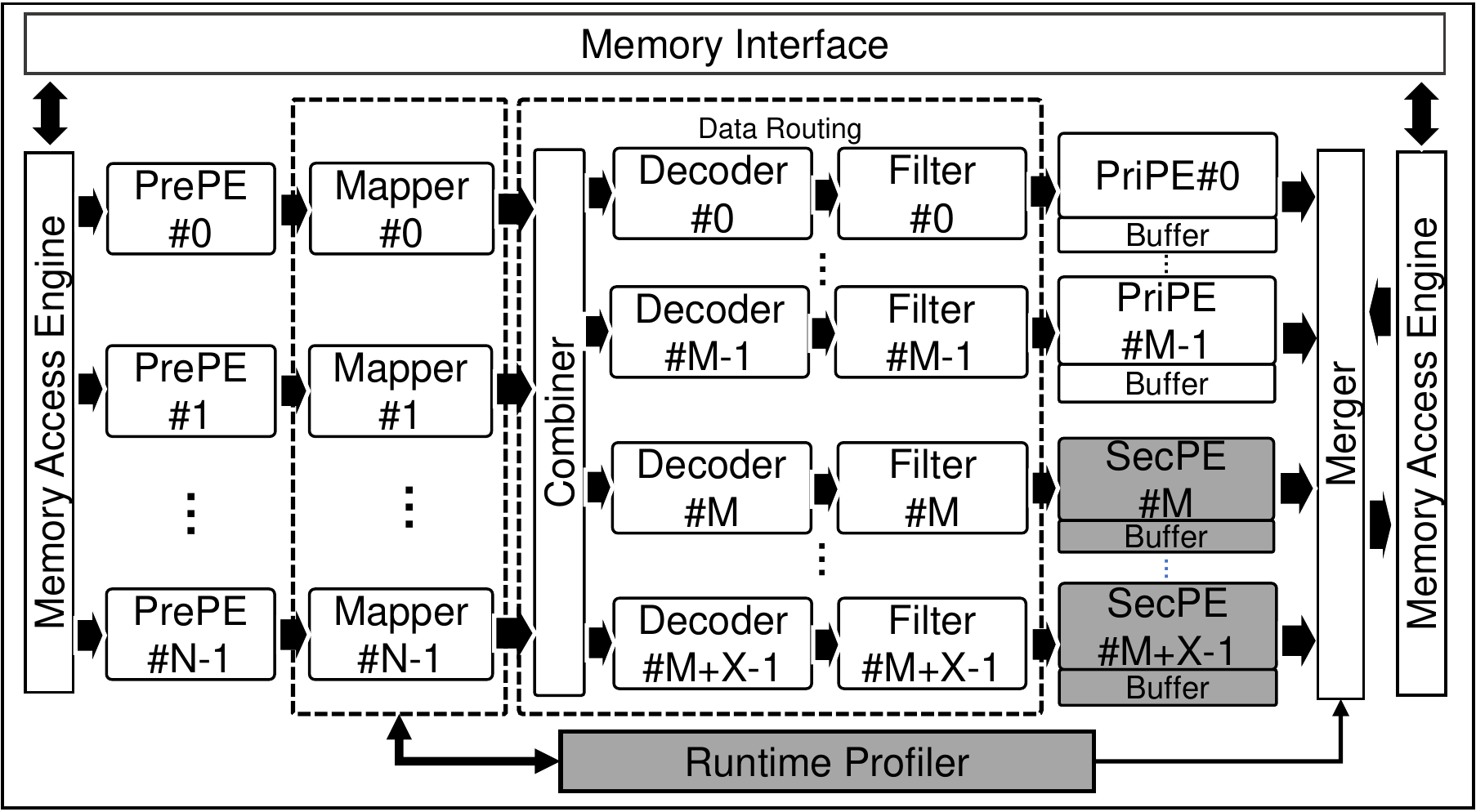}
    \caption{The skew-oblivious data routing architecture. Modules in solid grey color are enqueued and dequeued dynamically.} 
    \label{fig:template_arch}
\end{figure}

Fig.~\ref{fig:template_arch} shows the high-level architectural overview of our proposed skew-oblivious data routing architecture. 
It is composed of three types of PEs, which are the preprocessing PEs (PrePE), the primary PEs (PriPE) and the secondary PEs (SecPE). 
The logic of three kinds of PEs performs application-specific computations. % and needs to be specified by developers.
The $N$ PrePEs prepare the tuples with the format of $\langle dst, value \rangle$, where the $dst$ is the index of the buffered data and the value is to calculate with the buffered data. 
The $M$ PriPEs and the $X$ SecPEs are all accompanied with buffers and have the same logic for tuple processing.
They have been assigned unique identifies (IDs): $0$ to $M-1$ for PriPEs and $M$ to $M + X - 1$ for SecPEs. 
A PriPE processes a partial range of the input tuples, while a SecPE processes the same range of the tuples with the PriPE it is scheduled to.
Both the PriPEs and SecPEs become the designated PEs, and the data routing is responsible for dispatching the tuples to their designated PEs.
% PriPEs or SecPEs (both called destination PEs).

\subsection{Secondary PE scheduling}
{In order to schedule the $X$ SecPEs to help the overloaded PriPEs, the runtime profiler analyzes the workload distribution of PriPEs during the runtime to ascertain the overloaded PriPEs and then generates the SecPE scheduling plan for the mappers.}
According to the scheduling plan, the mappers redirect the tuples of the overloaded PriPEs to available SecPEs. %to alleviate the workload of overloaded PriPEs. 
By the end of the processing, the results of PriPEs and SecPEs are merged by the merger module according to the SecPE scheduling plan.
For non-decomposable applications such as data partitioning, PrePEs and SecPEs output results to their own memory space of the global memory.

A static SecPE scheduling plan cannot handle evolving data skew since the workload distribution varies during the runtime, as shown in Fig.~\ref{fig:workloaddistributionHISTO}.
In order to accommodate evolving datasets, the architecture reschedules SecPEs without interrupting the execution of PriPEs.
Once the runtime profiler ascertains workload distribution changed, it informs SecPEs and mappers and exits itself. 
The mappers will prevent the tuples from being routed to SecPEs.
The SecPEs exit the execution after all the tuples in the channels whose upstream is the data routing logic are consumed. 
The merger merges the {intermediate results in the global memory} with the results of SecPEs according to the SecPE scheduling plan.
After that, the CPU side enqueues the runtime profiler and SecPEs again; therefore, the SecPEs will be scheduled again according to the changed workload distribution.

\subsection{Design Details}
\subsubsection{Data Routing Logic}
The data routing dispatches $N$ tuples generated by the PrePEs to destination PEs per cycle.
We adopt the design from~\cite{chen2019fly} and simplify it into three modules for resource efficiency: the combiner, the decoder and the filter.
Since a tuple can be processed by any destination PE, the combiner gathers $N$ tuples together with their destination PE IDs and duplicates them for $M + X$ datapaths each owned by a destination PE. 
The decoder and the filter extract tuples to be processed by the current destination PE.
By comparing tuples' destination PE IDs with the current PE ID, the decoder generates an $N$ bits mask code, which marks the tuples to be processed.
It then outputs the positions and the number of tuples to be processed according to a preset table with the mask code as input. 
%The idea is that a tuple only has two statues (process or not to process) for a PE; hence, there are only $2^N$ processing possibilities of a set of $N$ tuples, which can be enumerated and stored in a table in advance.
The filter fetches the tuples to be processed according to the decoded information. 
Multiple concurrent kernels are used for the asynchronous execution of filters.
{This method resolves the run-time data dependency and enables high throughput~\cite{chen2019fly}}.

\subsubsection{{Mapper}}
\begin{figure}[t]
    \centering
    \includegraphics[width=1\linewidth]{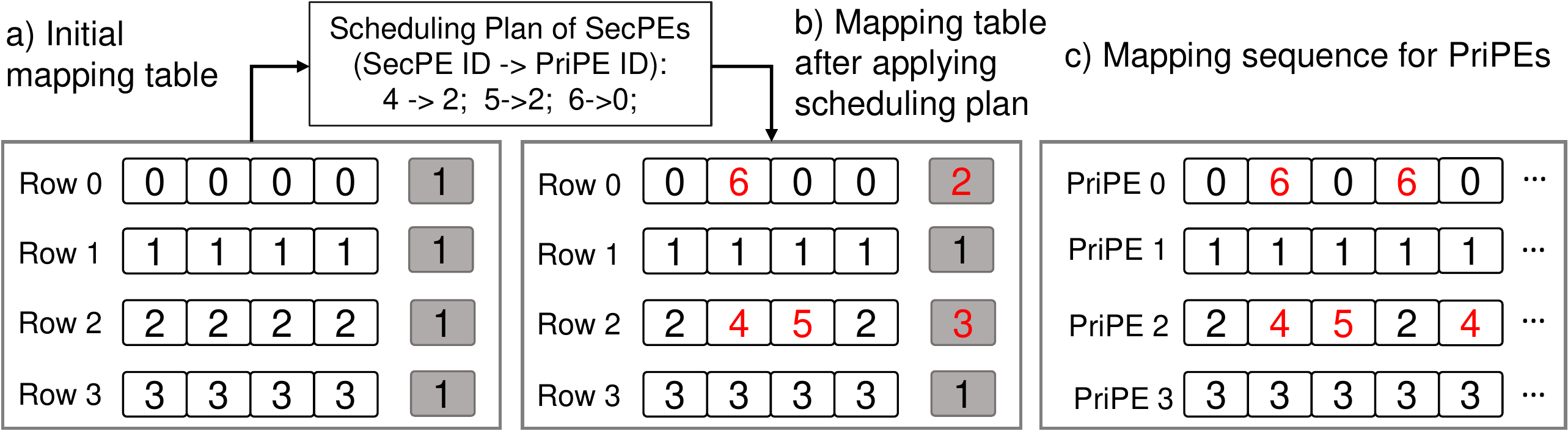}
    \caption{The mapping table updating and the mapping sequence for PriPEs, with four PriPEs and three SecPEs.} %{Light Unit}
    \label{fig:mapper}
\end{figure}

The mappers execute the SecPE scheduling plan by mapping the PriPE IDs to the designated SecPE IDs to redirect the workload of overloaded PriPEs to SecPEs. 
The map scheme is the scheduling plan of SecPEs.
%, is generated by the runtime profiler.
Each mapper maintains a two-dimensional mapping table with $M$ rows (for $M$ PriPEs) and $X + 1$ columns and a one-dimensional counter array with $M$ entries. 
The $X+1$ entries could accommodate a PriPE ID and all the schedulable SecPEs IDs ($X$).
The counter indicates the number of available PEs from the left side of the row and is initialized as one. 
The mapping table is initially filled with the PriPE ID and updated after the mapper receives the SecPE scheduling plan.
Fig.~\ref{fig:mapper}a shows an initial mapping table and the counter array with four PriPEs and three SecPEs.

\noindent
\textbf{Mapping table updating.}
The scheduling plan from the runtime profiler contains the array with ``SecPE ID $\to$ PriPE ID" pairs.
The mappers update only one pair to the mapping table per cycle for better timing. 
For a pair, the mappers write SecPE ID to the next position of the last available PE ID of the row by using the counter value as the write index and increase the counter by one. 
%And, the counter is increased by one.
Fig.~\ref{fig:mapper}b depicts the updated mapping table with the example SecPE scheduling plan. 
The SecPE 4 and SecPE 5 are written to the indices one and two of {row 2 which is} for PriPE 2 in two cycles, and the final counter is increased to three. 
The SecPE 6 is written to the second entry of {row 0 which is} for PriPE 0, with the counter increased by one.

\noindent
\textbf{Workload redirecting.}
The mappers redirect the destination PE ID by looking up the mapping table in a round-robin manner with the counter indicating the boundary.
Fig.~\ref{fig:mapper}c shows the mapping sequences. 
For example, the tuples with PE ID of 0 will go to SecPE 6 in odd cycles, and the tuples with PE ID of 2 will go to PriPE 2, SecPE 4, and SecPE 5 in a round-robin manner. 
%In the end, the workloads of over-loaded PriPE are amortized by the attached SecPEs.

\subsubsection{Runtime Profiler}
The runtime profiler performs two operations: 1) generating the SecPE scheduling plan by monitoring the workload distribution among PriPEs; 2) informing the system to reschedule SecPEs if the workload distribution has changed. %system throughput has downgraded to a predefined threshold.

\begin{figure}[h]
    \centering
    \includegraphics[width=1\linewidth]{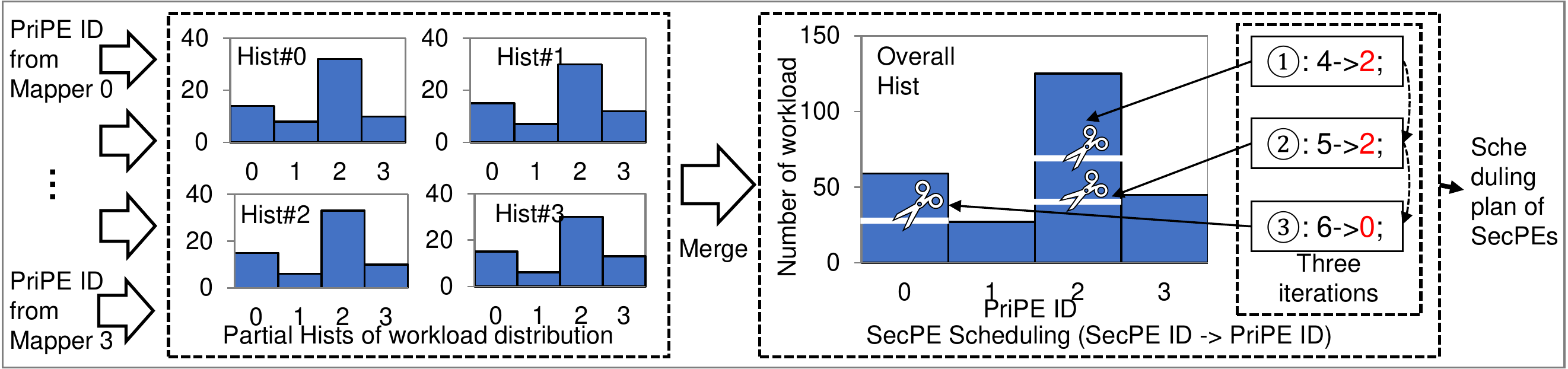}
    \caption{The generation process of SecPE scheduling plan, with four PrePEs, four PriPEs and three SecPEs.} %{Light Unit}
    \label{fig:profiler}
\end{figure}

\noindent
\textbf{SecPE Scheduling plan generation.} 
The runtime profiler receives $N$ PriPE IDs from the mappers in one cycle with $N$ independent hist instances which count the number of tuples processed by each PriPE. 
After a certain number of profiling cycles, it terminates the workload counting and merges the $N$ partial results into a global histogram which indicates the workload distribution among PriPEs. Fig.~\ref{fig:profiler} shows an example with the profiling cycles of 256.
The runtime profiler assigns a SecPE to the PriPE whose workload is maximal and recalculates the workload distribution with assuming the original workload is evenly shared with the attached SecPEs. This process is repeated until all SecPEs are scheduled.
In the example shown in Fig.~\ref{fig:profiler}, the PriPE 2 has the maximal workload for the first two iterations, and hence its workload is divided to one-third because of the involvement of 2 SecPEs. 
The final scheduling plan of $X$ SecPEs is recorded through an array with $X$ entries and transferred to the mappers and the merger. 
Since scheduling plan generation is not on the critical path of the overall execution pipeline, we make it serially executed to reduce the resource consumption.

\noindent
\textbf{Workload distribution monitoring.} 
The runtime profiler also monitors the system throughput to determine SecPEs rescheduling. 
On the one hand, it keeps receiving the number of tuples processed from the mappers. 
On the other hand, it maintains a local counter as a clock tick, which increases one per cycle.
The system throughput is calculated by the incremental number of processed tuples in a certain number of clock ticks. 
Once the system throughput downgrades to the predefined threshold, which means the workload distribution has changed, it informs the mappers and SecPEs and exits itself. 
The predefined threshold can be set to zero to stop the SecPE rescheduling if the time interval of workload distribution changing is smaller than kernel dequeueing and enqueueing overhead.
%It also has a mechanism to avoid frequent rescheduling. 

\begin{figure*}[t]
    \centering
    \includegraphics[width=0.8\linewidth]{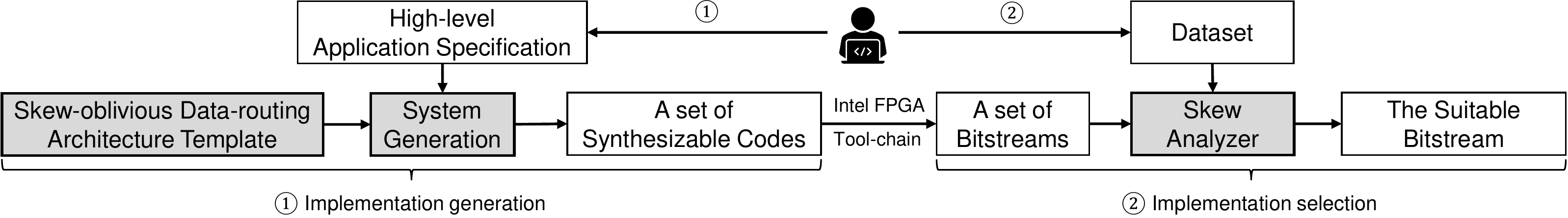}
    \caption{The workflow of Ditto.}
    \label{fig:framework_overview}
    \vspace{-4mm}
\end{figure*}

\subsubsection{Other Optimizations}
The architecture also adopts other HLS optimizations~\cite{cong2018automated}. Firstly, the memory access engine coalesces memory requests and accesses the global memory in a burst manner for high memory bandwidth utilization. Secondly, the modules of the design are connected with channels and pipelined for high efficiency. 
%with synchronization among kernels and exit condition of kernels carefully handled. 
%synchronization among kernels and exit condition of kernels~\cite{SDKdocument2018} are \textcolor{blue}{carefully optimized to achieve an efficient pipeline (carefully optimized is not the optimization we have done, can we simply say pipelined?)}. 
%Lastly, as the SecPEs are reschedulable at runtime, the interfacing logic between SecPEs and the rest of the design is \textcolor{blue}{carefully implemented (is careful implementation also an optimization?)} for correct results and high efficiency. 

\section{The Ditto framework}\label{sec:frameworkoverview}
In this section, we introduce our Ditto framework, which is integrated with the skew-oblivious data routing architecture to reduce manual efforts of skew handling for data-intensive applications. 

\subsection{Overview of Ditto}\label{sec:dittooverview}
As shown in Fig.~\ref{fig:framework_overview}, the workflow with Ditto is composed of two phases: implementation generation and implementation selection.
First, the developers describe the high-level application specification with the provided programming interface and set the parameters, including the data type of the tuples and data width of the memory interface. 
Then, together with the skew-oblivious data routing architecture template, the system automatically generates a set of synthesizable codes which have different numbers of SecPEs. 
Note that the system is currently built with Intel's OpenCL tool-chain~\cite{SDKdocument2018} to generate the bitstreams, but it can be migrated to the Xilinx OpenCL tool-chain as well.
Next, given the dataset by the developer, the system evaluates the data skew in the dataset to select the most suitable implementation that could save the BRAM usage without significantly compromising the performance.

\subsection{Programming Interface}\label{sec:dittotemplate}
We implement the proposed skew-oblivious data routing architecture as a hardware architecture template in the form of an HLS library.
%The numbers of PrePE, PriPE and SecPEs, the profiling cycles of the workload distribution, and the data type of tuples are parameterized in the template to ease the system generation. 
The template provides inline functions to specify the processing logic of PEs.
Listing~\ref{listing:example_histbuilding} demonstrates the high-level specification of histogram building (in the red rectangles).
The PrePEs read the tuples from the global memory by the provided channel (line 4). 
The developers then assign the rule of data routing that the destination PE ID of the tuple is formed by the four least significant bits of the key (line 5). 
Tuples, together with their destination PE IDs, are written to the downstream channel for data routing (line 6). 
PriPEs and SecPEs receive the tuples from the data routing (line 13) and build the partial histograms in the local buffer (lines 14 to 15).
%The system takes the rest to merge histograms of PrePEs and SecPEs and outputting them to the global memory. 
%For non-decomposable applications such as data partitioning, the developers should choose PrePEs and SecPEs to output to their own memory space of the global memory.

\begin{minipage}[b]{1\linewidth}
\begin{lstlisting}[escapechar=!,language=C++, caption=High-level specification of histogram building., label={listing:example_histbuilding}]
__attribute__(task)
__kernel void PrePE (cl_channel channel){
	while (true){
	!\tikzmark{a}!  tuple = channel.read();
		dst = tuple.key & 0xf; 
		channel.write(tuple, dst);!\tikzmark{b}!
	}
	//Other logic provided by the template
}
__attribute__(task)
__kernel void PriPEandSecPE (cl_channel channel){
	while (true){
	!\tikzmark{c}!  tuple = channel.read(); 
		idx = HASH (tuple.key);
		hist[idx] ++;             !\tikzmark{d}!
	}
	//Other logic provided by the template
}
\end{lstlisting}

\begin{tikzpicture}[remember picture,overlay]
\draw[red,rounded corners]
  ([shift={(3pt,1.25ex)}]pic cs:a) 
    rectangle 
  ([shift={(3pt,-0.65ex)}]pic cs:b);
\end{tikzpicture}

\begin{tikzpicture}[remember picture,overlay]
\draw[red,rounded corners]
  ([shift={(3pt,1.25ex)}]pic cs:c) 
    rectangle 
  ([shift={(3pt,-0.65ex)}]pic cs:d);
\end{tikzpicture}
\vspace{-10mm}
\end{minipage}

\subsection{System Generation}\label{sec:sysgeneration}
With the high-level specification of the application from developers, the system firstly tunes the numbers of PriPEs and PrePEs of the template to balance the pipeline and then generates a set of synthesizable codes with varying the number of SecPEs.

The numbers of PriPEs and PrePEs are tuned to form a balanced pipeline and satiate the memory bandwidth of the platform. 
The logic programmed by developers will be synthesized by the HLS tool to get the estimated initiation interval (II)~\cite{SDKdocument2018} of PrePEs and PriPEs: $II_{\text{PrePE}} $ and $ II_{\text{PriPE}}$, respectively.
Suppose the data width of the memory interface is $W_{\text{mem}}$, and the data width of the tuple is ${W_{\text{tuple}}}$.
The numbers of PrePEs ($N_{\text{PrePE}}$) and PriPEs ($N_{\text{PriPE}}$) are calculated by Equation~\ref{eq:num_PE}.
Subsequently, the system tunes data routing by invoking different sets of codes that have different routing entries.

\begin{equation}\label{eq:num_PE}
    \frac{N_{\text{PrePE}}}{II_{\text{PrePE}}} = \frac{N_{\text{PriPE}}}{II_{\text{PriPE}}} = \frac{W_{\text{mem}}}{W_{\text{tuple}}}
\end{equation}

The system then generates $M$ sets of codes with the number of SecPEs ranging from 0 to $M-1$, which trades off the capacity of skew handling against the BRAM usage. 
Assuming the available BRAM size for buffering data is $C$, the maximal size of buffered distinctive data is $\frac{M}{(M + X)} \times C$ with $X$ SecPEs.
When there is no SecPEs ($X=0$), the implementation could use all the capacity, $C$, to buffer distinctive data.
The upper bound of $X$ is $M-1$ since the implementation with $M-1$ SecPEs could handle the worst case where all data go to the same PriPE.
In other words, the system could buffer $\frac{C}{2}$ distinctive data at least.

\subsection{Skew Analyzer}\label{sec:analyzer}
Given the generated hardware implementations, the skew analyzer chooses the implementation with a suitable number of SecPEs according to the skew level of the dataset. 

For offline processing, the skew analyzer randomly samples a certain number of data of the dataset to analyze the workload distribution among PriPEs.
The suitable number of SecPEs, $X$, is calculated by Equation~\ref{eq:imp_select}.
The ${\mbox{workload}_{\text{PriPE}_{i}}}$ indicates the sampled workload of the \text{PriPE} ${i}$. $M$ is the total number of PriPEs, and $T$ is the tolerance factor indicating the performance compromise in terms of percentages. 
This equation guarantees that the workload of a PriPE after evenly shared with SecPEs is less than that of the uniformly distributed dataset where every PriPE has the same workload; hence, the processing is not bottlenecked by any PriPE.
In addition, the maximal $X$ is $M-1$.
%The sampling process is executed multiple times to ensure it covers the data skew.
The developer could also choose the number of SecPEs manually to set a required buffer size for distinctive data, but this will override the build-in implementation selection mechanism.

\begin{equation}\label{eq:imp_select}
  X = \sum_{i=1}^{M}\vast\lceil{\left|\frac{M \times \text{workload}_{\text{PriPE}_{i}}}{\sum_{i=1}^{M} {\text{workload}_{\text{PriPE}_{i}}}} - T \right|}\vast\rceil - M
\end{equation}

For online processing, as the dataset is a prior information, the skew analyzer currently chooses the implementation with the maximal number of SecPEs, $M-1$, to accommodate any level of data skew. 
There are a number of works on predicting the future input of stream processing~\cite{predicateinput}, which can be explored for choosing an implementation that saves more BRAM usage for online processing.
\section{Evaluation}\label{sec:evaluation}

%To demonstrate the effectiveness and robustness of Ditto, 
We evaluate Ditto on five commonly used applications. 
Specifically, we compare Ditto with the state-of-the-art designs in Section~\ref{sec:evauniform}. 
As those implementations are designed mostly for uniform datasets, we use inputs with uniform distributions for a fair comparison. 
In Section~\ref{sec:evastaticdataskew}, we evaluate Ditto with inputs with varying data skew while studying the effectiveness of our proposed technique. 
In Section~\ref{sec:evaevolvingskew}, we use a case of evolving dataset to demonstrate the capability of Ditto in adapting to dynamic data skew.

\subsection{Experimental Setup}
\subsubsection{Hardware Platform}\label{sec:hardware}
Experiments are conducted on Intel's PCIe Programmable Acceleration Card (PAC) which is featured with an Arria 10 GX FPGA and 2$\times$4GB DDR4 memory. 
The FPGA device has 1,150K logic elements, 65.7 Mb of on-chip memory and 3,036 digital signal processing (DSP) blocks. 
The development tool is the Intel FPGA SDK for OpenCL, version 17.1.1. 

\subsubsection{Applications}
Five representative datacenter applications from database field, graph processing and in network processing are evaluated, which are  histogram building~\textbf{(HISTO)}, data partitioning~\textbf{(DP)}, pagerank~\textbf{(PR)}, hyperloglog~\textbf{(HLL)}, and heavy hitter detection~\textbf{(HHD)}, detailed descriptions shown in Table~\ref{table:applications}. 

\begin{table}[t]
\centering
\caption{Application details.}
\label{table:applications}
\resizebox{1\linewidth}{!}{%
\begin{tabular}{lll}
\toprule
\multicolumn{1}{l}{App.}   & Description & Algorithm details \\ 
\midrule
\multicolumn{1}{l|}{HISTO} &Represents the distribution of numerical data & with equi-width histograms\\ 
\multicolumn{1}{l|}{DP}   &  Separates a big dataset into many chunks & with radix hash function\\ 
\multicolumn{1}{l|}{PR}  & Scores the importance of websites by links & with fixed-point data type\\ 
\multicolumn{1}{l|}{HLL} & Estimates the cardinality of the big datasets & with murmur3 hash function\\ 
\multicolumn{1}{l|}{HHD} & Detects heavy hitters in the data streams & with the count-min sketch\\ 
\bottomrule
\end{tabular}
}
\end{table}

\subsection{Comparison with State-of-the-art Designs on Uniform Datasets}\label{sec:evauniform}
Table~\ref{table:compare} shows the comparison between system-generated implementations with state-of-the-art designs.
We reproduce the results from the open-source implementations (marked as Reproduced) while collecting the results from original papers for those not opensourced (marked as Original). 
It is noteworthy that PR from Chen et al.~\cite{chen2019fly} and HISTO from Jiang et al.~\cite{jiang2020boyi} have around 800 and 200 lines of kernel code, respectively, while Ditto requires only 22 and 6 lines, respectively.
We use the datasets described in corresponding papers, which are mostly uniformly distributed except the dataset of HHD has half of the tuples with the same key. 
The bandwidth is normalized for a fair comparison except for Kara et al.~\cite{kara2017fpga} as their platform has different random memory access performance. 
%The metric of throughput is the number of tuples processed per second. 
%The BRAM usage efficiency is obtained by architecture analysis. 
%For in-network processing applications: HLL and HHD, our system currently uses the memory interface to simulate the network interface. 

The results show that the implementations with Ditto outperform most of the existing implementations.
Compared with HLS-based ones, data routing resolves the run-time data dependency of DP~\cite{Wang2017} and omits the CPU intervention of HISTO~\cite{jiang2020boyi}.
Performance of PR is the same as Chen et al.~\cite{chen2019fly} since both implementations adopt data routing and directed graphs have near balanced workload distribution.
Compared with the RTL-based ones, our HHD outperforms work~\cite{heavyhitter} which only has one PE.
Our HLL has similar performance with work~\cite{hyperloglog} as both designs fully utilize the available bandwidth.
The BRAM usage per PE of our implementation is significantly reduced because of the avoidance of the data replication in buffers and the reduction can reach up to 32$\times$. 
{This improvement delivers critical benefits for the above data-intensive applications. Specifically, HISTO achieves a finer-grained distribution, HLL obtains more accurate estimation, and DP reaches a higher fan-out.}
%As a sanity check, we compare the lines of code in the state-of-the-art implementation to the implementation based on Ditto.

\begin{table}[t]
\centering
\caption{Ditto compared to the state-of-the-art designs in terms of programming language (P.L.), throughput (Thro.), and BRAM usage saving per PE (B.U.Saving).}
\label{table:compare}
\resizebox{0.96\linewidth}{!}{%
\begin{tabular}{@{}llcccc@{}}
\toprule
\multicolumn{1}{l}{App.} & Existing Works &\multicolumn{1}{l}{Source}   & \multicolumn{1}{l}{P.L.} & \multicolumn{1}{l}{Thro.} & \multicolumn{1}{l}{B.U.Saving} \\ \midrule
\multicolumn{1}{l|}{HISTO} & Jiang et al.~\cite{jiang2020boyi}              &Reproduced   & HLS & \textbf{1.2 $\times$} & \textbf{32 $\times$} \\ \hline
\multicolumn{1}{l|}{\multirow{2}{*}{DP}} & Wang et al.~\cite{Wang2017}      &Original    & HLS & \textbf{2.4 $\times$} & \textbf{16 $\times$} \\ \cline{2-6} 
\multicolumn{1}{l|}{} & Kara et al.~\cite{kara2017fpga}                     &Original    & RTL & \textbf{1.2 $\times$} & \textbf{8 $\times$} \\ \hline
\multicolumn{1}{l|}{\multirow{2}{*}{PR}} & Chen et al.~\cite{chen2019fly}   &Reproduced    & HLS & \textbf{1.0 $\times$} & \textbf{1 $\times$} \\ \cline{2-6} 
\multicolumn{1}{l|}{} & Zhou et al.~\cite{zhou2019hitgraph}                 &Original    & RTL & \textbf{1.8 $\times$} & \textbf{1 $\times$} \\ \hline
\multicolumn{1}{l|}{HLL} & Kulkami et al.~\cite{hyperloglog}                &Original    & RTL & \textbf{0.9 $\times$} & \textbf{10 $\times$} \\ \hline 
\multicolumn{1}{l|}{HHD} & Tong et al.~\cite{heavyhitter}                   &Original    & RTL & \textbf{1.6 $\times$} & \textbf{1 $\times$} \\ \bottomrule
\end{tabular}%
}
\end{table}

\subsection{Evaluation on Static Data Skew}\label{sec:evastaticdataskew}
\subsubsection{On Zipf dataset}
\begin{figure}[h]
    \centering
    \includegraphics[width=1\linewidth]{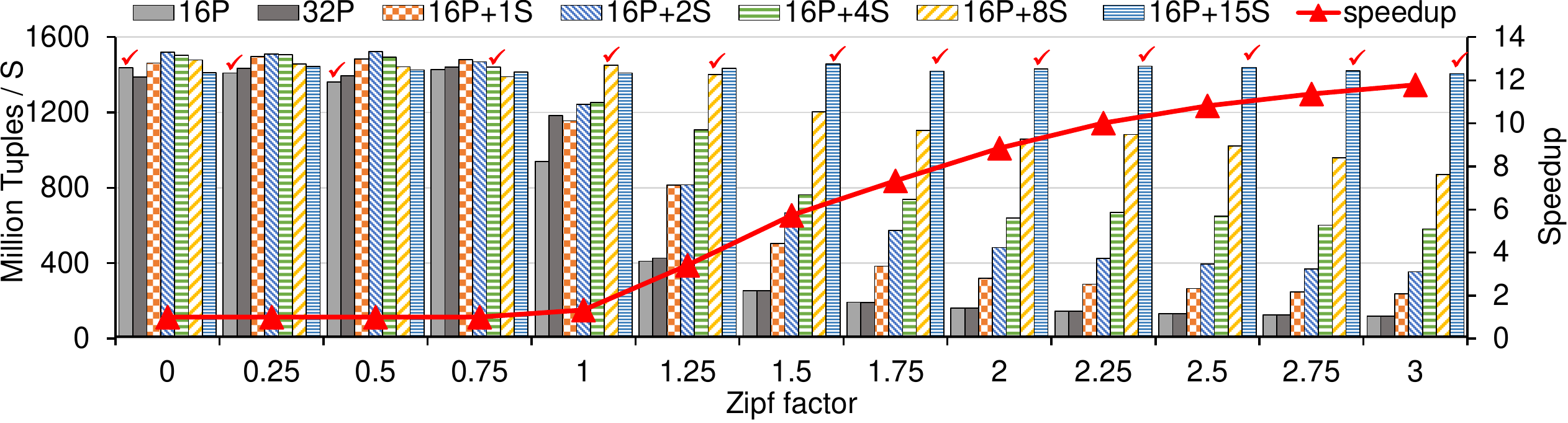}
    \caption{The throughput of HLL implementations with different number of SecPEs over Zipf distributions.} %{Light Unit}
    \label{fig:hll-skew}
    \vspace{-1mm}
\end{figure}

We evaluate the robustness of implementations with different numbers of SecPEs under the Zipf distribution~\cite{Balkesen2013} as well as the implementation selection of Ditto. 
Besides, we compare with the solution that simply increases the number of PriPEs.
With 8-byte tuples, the system sets the number of PriPEs to 16 on our platform.
\blue{
We let Ditto sample only 0.1\% of the total dataset  (256 *
100 data points sampled), which takes 0.047ms with a single thread of Xeon Platinum 8180 CPU.
}
%The tuples are of 8-byte, and the theoretical throughput is 1600 million tuples per second under 200MHz. 
As the trends observed are similar for the five applications, we take the throughput of HLL as an instance, as shown in Fig.~\ref{fig:hll-skew}.
The ticks indicate the implementations chosen by the system with \textit{T} of 0.01 and the red line stands for its speedup over the baseline which only has 16 PriPEs (16P). 
%\todo[inline]{What is this 16P?}
The resource utilization, together with frequency, is shown in Table~\ref{table:Resource-utilization}. 
It is noteworthy the runtime profiler module only costs 6\% logic and 8\% DSPs.

Based on the results, we have the following {observations}. 
Firstly, implementations with more SecPEs are more robust to heavier data skew and deliver larger speedup (up to 12$\times$ speedup on extreme data skew). 
{The ``16P+15S" implementation is oblivious to any skew and indicates that the upper bound of $X$ is $M - 1$.}
Secondly, increasing the number of PriPEs (32P) could not help the performance as the PE overloading is not solved.
Thirdly, Ditto could select a suitable implementation that minimizes the BRAM usage without compromising performance.
Lastly, the resource consumption is growing up with more SecPEs but not proportional due to the static resource consumption of the built-in shell~\cite{SDKdocument2018}. 
%\textcolor{blue}{Still, the logic resource is not the bottleneck, indicating exchanging the logic for other benefits such as BRAM usage efficiency is feasible.}
%as data intensive-applications require less computation.

\begin{table}[b]
\vspace{1mm}
\caption{The resource utilization and frequency of HLL implementations with different number of SecPEs.}
\label{table:Resource-utilization}
\setlength{\tabcolsep}{8pt} % Default value: 6pt
\renewcommand{\arraystretch}{0.6} % Default value: 1
\resizebox{1\linewidth}{!}{%
\begin{tabular}{lllllll}
\toprule
\multicolumn{1}{l}{Implem.}   & Frequency & RAM  & Logic & DSP   \\ \midrule
\multicolumn{1}{l|}{16P}    & 246 MHz    & 597 (22\%) & 163,934  (38\%) & 403 (27\%) \\
\multicolumn{1}{l|}{32P}    & 191 MHz    & 1,868 (69\%) & 230,838 (60\%) & 729 (48\%) \\
\multicolumn{1}{l|}{16P+1S}    & 202 MHz    & 908 (33\%) & 184,826 (43\%) & 409 (27\%)     \\
\multicolumn{1}{l|}{16P+2S}   & 180 MHz    & 1,021 (38\%) & 203,083 (48\%) & 575 (38\%) \\ 
\multicolumn{1}{l|}{16P+4S}    & 192 MHz    & 1,309  (48\%) & 212,856 (50\%) & 587 (39\%)  \\ 
\multicolumn{1}{l|}{16P+8S}    & 196 MHz    & 1,374 (51\%) & 281,667 (66\%) & 616 (41\%)  \\ 
\multicolumn{1}{l|}{16P+15S}   & 188 MHz    & 2,129 (78\%) & 230,095 (54\%) & 658 (43\%)  \\ 
\bottomrule
\end{tabular}%
}
\end{table}

\subsubsection{On irregular graph structure}
\begin{figure}[t]
    \centering
    \includegraphics[width=1\linewidth,inner]{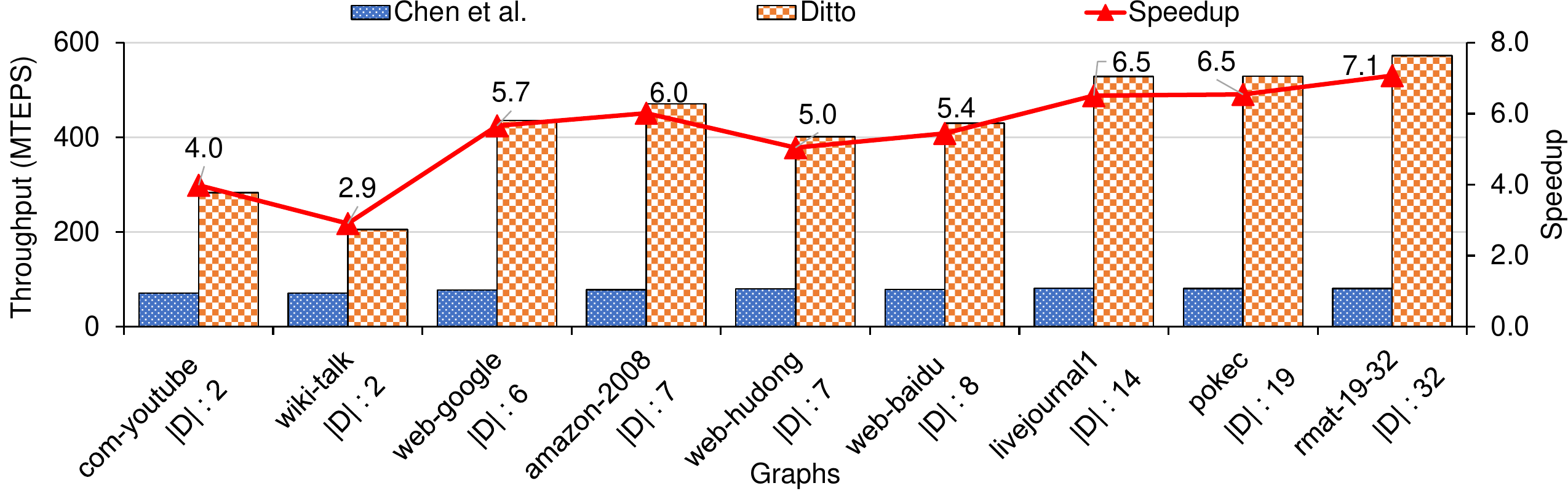}
    \caption{The throughput comparison of PR on undirected graphs.} %{Light Unit}
    \label{fig:graphskew}
    \vspace{-1mm}
\end{figure}

We then evaluate the generated PR implementation of Ditto on undirected graphs.
Fig.~\ref{fig:graphskew} shows the performance comparison with the state-of-the-art design~\cite{chen2019fly} on public graphs~\cite{graphdataset} and synthetic graphs~\cite{chen2019fly}, where the graphs shown in the x-axis are in ascending order by their degrees. 
%The system selected implementation is ``16P+10S" with 8-byte tuples. 
The million traversed edges per second (MTEPS) is used as the throughput metric.
The results show that the throughput of Ditto is significantly improved and can be up to 7$\times$ of the existing solution. 
In addition, the speedup grows up with a larger graph degree since more edges updating the same vertex causes more severe data skew.

\subsection{Evaluation on Evolving Data Skew}\label{sec:evaevolvingskew}
\begin{figure}[h]
    \centering
    \includegraphics[width=1\linewidth, inner]{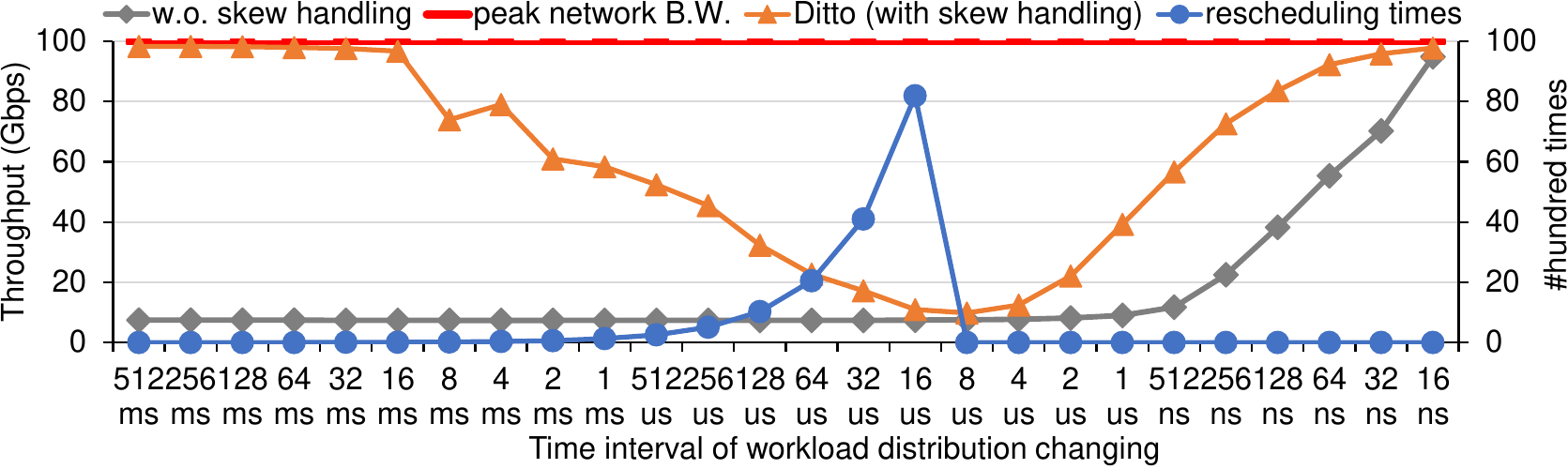}
    \caption{The throughput comparison of HISTO with varying the time interval of changing workload distribution.} %{Light Unit}
    \label{fig:varyingskew}
    \vspace{-1mm}
\end{figure}

We evaluate the robustness of Ditto on evolving data skew by emulating an online processing  scenario with HISTO with 16P+15S and 8-byte tuples. 
We set the Zipf factor to three and vary the seeds of the dataset generator for generating different workload distributions.
The memory interface is used to simulate the 100 Gbps network interface.
Fig.~\ref{fig:varyingskew} shows the throughput of HISTO with varying the time interval of changing workload distribution. 

The results show the following highlights. 
First of all, Ditto consistently achieves better performance than the baseline which does not have skew handling.
Secondly, the throughput is able to satiate the network bandwidth when the time interval is larger than 16 ms; whereas, it drops significantly for intervals between {16 ms and 64 ns} because the overhead of SecPE rescheduling leads SecPEs underutilized. 
Lastly, the throughput increases to meet the bandwidth again since the internal channels could accommodate short-term skew distribution variances; meanwhile, the system stops rescheduling SecPEs as the interval is smaller than the rescheduling overhead.

%\vspace*{-1mm}
\section{Related Work}\label{sec:relatedwork}
Load-balancing problem of FPGA-based designs has been studied extensively.
Ramanathan et al.~\cite{workstealing} studied HLS-based work-stealing with K-means algorithm as a case study, and the OpenCL atomic operation is used for synchronization among PEs.
Later, Yan et al.~\cite{yan2019constructing} improved their performance by replacing the work-stealing with a round-robin work distributor. 
Geng et al.~\cite{awbgcn} proposed an HDL-based runtime workload balancing method for graph convolutional networks.
%A common pattern of the problems the above solutions addressed is that the computation is the bottleneck of the pipeline; whereas, 
Those studies mainly focus on the problems where the computation is the key performance bottleneck. In contrast, data-intensive applications involve only lightweight computation and require multiple workloads dispatched in one cycle. 
Likewise, software-based flexible skew handling methods for stream processing~\cite{cpuskewhandling} are too heavy to fit into cycle-level requirements.    
%-- add software solution and their failure reasons. 
%/add skew handling in software like database, and why they cannot be used here... 2-3 sentences are sufficent.

%Meanwhile, there have been a number of works for improving the performance of HLS-based designs. 
There are quite some studies improving the performance of HLS-based designs with static workload dispatching to PEs.
Cong et al. proposed~\cite{cong2018automated} a composable microarchitecture to reduce the design space and further integrated it into a framework to automate the entire accelerator generation process. 
Thomas et al. introduced Fleet~\cite{thomas2020fleet}, which duplicates the user's processing logic to feed the units with separate streams and drain their outputs.
%However, two works statically dispatch workloads to PEs; hence, they are more suitable for applications expressing parallelism by data partitioning.
Wang et al. proposed Melisa~\cite{wang2016melia} to extend the MapReduce framework to OpenCL-based FPGAs; nevertheless, the optimizations of underlying architecture are largely overlooked.
Cong et al. also presented buffer restructuring approaches~\cite{cong2017bandwidth} to optimize the bandwidth utilization with HLS. 
In this work, we have combined them into the memory access engine.

%ST-Accel~\cite{ST-Accel} proposed by Ruan et al. addresses the inefficiency of applying OpenCL to the streaming workload by enabling the host/FPGA communication during kernel execution. 
%This study is orthogonal with Ditto since we focus on accelerator design inside of FPGA. 
%Li et al. augured current HLS tools could only handle regular applications with static analysis at compile time, hence, they propose a aggressive pipeline architecture to assist HLS tools handle irregular applications with run-time dependency.  
%There are also many works arguing the limitations of HLS tools, hence propose their own frameworks.  
%Chen et al. proposed a framework~\cite{chen2018architectural} for accelerating dynamic parallel algorithms on FPGA. 
\section{Conclusion}\label{sec:conclusion}
%\st{While HLS-based data routing architecture enables more flexibility and more potential of BRAM usage for data-intensive applications, the workload imbalance problem caused by skew datasets remains unsolved.}
In this paper, we propose an HLS-based skew-oblivious data routing architecture that solves the workload imbalance problem caused by skew datasets of the original data routing. 
In order to ease the skew handling for a class of data-intensive applications, we further propose a framework, \textit{Ditto}, which only requires high-level application specification and generates an implementation that saves BRAM usage and handles data skew.
%hardware implementations with different capacities on skew handling and select the most suitable implementation for the given dataset. 
The five representative applications implemented with our framework demonstrate robust performance on skew datasets and outperform existing works in terms of both throughput and BRAM usage. 
Our study also sheds light that HLS could accomplish intricate hardware designs with careful optimizations and deliver HDL-comparable performance.

\section*{Acknowledgement}
\blue{
We thank Intel Labs Academic Compute Environment for remote hardware access.
This work is supported by MoE AcRF Tier 1 grant (T1 251RES1824), Tier 2 grant (MOE2017-T2-1-122) in Singapore, and also partially supported by the National Research Foundation, Prime Minister's Office, Singapore under its Campus for Research Excellence and Technological Enterprise (CREATE) programme.
}

%^\AtBeginBibliography{\small}
\bibliographystyle{IEEEtran}
\bibliography{IEEEabrv,reference_dac}

\end{document}